\documentclass[prb,superscriptaddress ,twocolumn, aps,fleqn,floatfix]{revtex4-1}

\usepackage{amsmath}
\usepackage{amstext}
\usepackage{amsfonts}
\usepackage{graphicx}
\usepackage{draftcopy}
\usepackage{flafter}
\usepackage[below]{placeins}
\usepackage{float}
\usepackage{tabularx}

\graphicspath{{Figs/}}

\begin{document}

\title{Magnetization Noise Induced Collapse and Revival of Rabi Oscillations in circuit QED}

\author{Amrit De}
\affiliation{Department of Physics, University of Wisconsin - Madison, WI 53706}
\affiliation{Department of Physics and Astronomy, University of California - Riverside, CA 92521}
\author{Robert Joynt}
\affiliation{Department of Physics, University of Wisconsin - Madison, WI 53706}


\begin{abstract}
We use a quasi Hamiltonian formalism to describe the dissipative dynamics of a circuit QED qubit that is affected by several fluctuating two level systems with a $1/f$ noise power spectrum. The qubit-resonator interactions are described by the Jaynes Cummings model. We argue that the presence of pure dephasing noise in such a qubit-resonator system will also induce an energy relaxation mechanism via a fluctuating dipole coupling term. This random modulation of the coupling is seen to lead to rich physical behavior. For non-Markovian noise, the coupling can either worsen or alleviate decoherence depending on the initial conditions. The magnetization noise leads to behavior resembling the collapse and revival of Rabi oscillations. For a broad distribution of noise couplings, the frequency of these oscillations depends on the mean noise strength. We describe this behavior semi-analytically and find it to be independent of the number of fluctuators. This phenomenon could be used as an {\it in situ} probe of the noise characteristics.
\end{abstract}

\maketitle

\section{Introduction}\label{sec.intro}

In the area of quantum information processing(QIP), cavity quantum electrodynamics(QED) has received much attention as it allows the inter conversion between matter based qubits and photonic qubits. There are several promising cavity QED schemes -- ion traps\cite{Leibfried2003}, neutral atoms\cite{Thompson1992}, Rydberg atoms \cite{Raimond2001}, quantum dots in photonic crystals\cite{Hennessy2007}. However superconducting qubits coupled to transmission line resonators (or LC-circuit oscillators) \cite{Devoret1989,Schoelkopf2008,Martinis2009,Girvin2009} appear to be extremely promising as they offers two substantial advantages. Firstly, existing fabrication and lithographic techniques can be used to make a large number of scalable superconducting qubits, which act as macroscopic quantum objects. Secondly, the smaller mode volume of the resonators and the larger physical size of the qubits, allows strong coupling between the cavity mode and the qubit, enabling efficient manipulation of interactions at the single photon level.

In cavity QED systems and particularly in the case of circuit QED schemes, there are three main mechanisms that contribute to the qubit's decoherence. They are: decay into photon modes other than the cavity mode, relaxation due to vacuum fluctuations and dissipation due to electrical noise. The first one is negligible for transmission line resonators largely due to their 1-D nature. The relaxation due to vacuum fluctuations can be reduced by increasing the level detuning\cite{Blais2004}. The third category can be further divided into extrinsic and intrinsic sources of noise. In case of SC flux/phase qubits, extrinsic noise sources, such as ones due to the fluctuations in the external circuitry can, for example, be fixed by increasing the impedance of the current source providing the flux bias \cite{Makhlin2001}. On the other hand intrinsic noise of the $1/f$ type \cite{Yoshihara2006,Bialczak2007}, is often extremely difficult to deal with. However, it has been shown that two level fluctuators contributing to $1/f$ noise can be minimized by suitably engineering the Joshephson tunnel junction \cite{Martinis2005,Oh2006}.

While $1/f$ type magnetic flux noise was observed in SQUIDs quite some time ago\cite{Koch1983,Wellstood1987} its origins remained somewhat unexplained. Recent interest in quantum computing has however renewed interest in this field\cite{Yoshihara2006,Bialczak2007}. A number of models have been suggested to explain this phenomenon. Examples include a model proposed by Koch {\it et al} \cite{Koch2007} where unpaired and non-interacting electrons randomly hop between traps with fixed but random spin orientations. The trap energies in their model have $1/f$ distribution. A dangling bond model\cite{deSousa2007} for $1/f$ noise has been proposed where electrons flip their spins due to their interaction with tunneling two-level-systems(TLS).

\begin{figure}
\centering
\includegraphics[width=0.8\columnwidth]{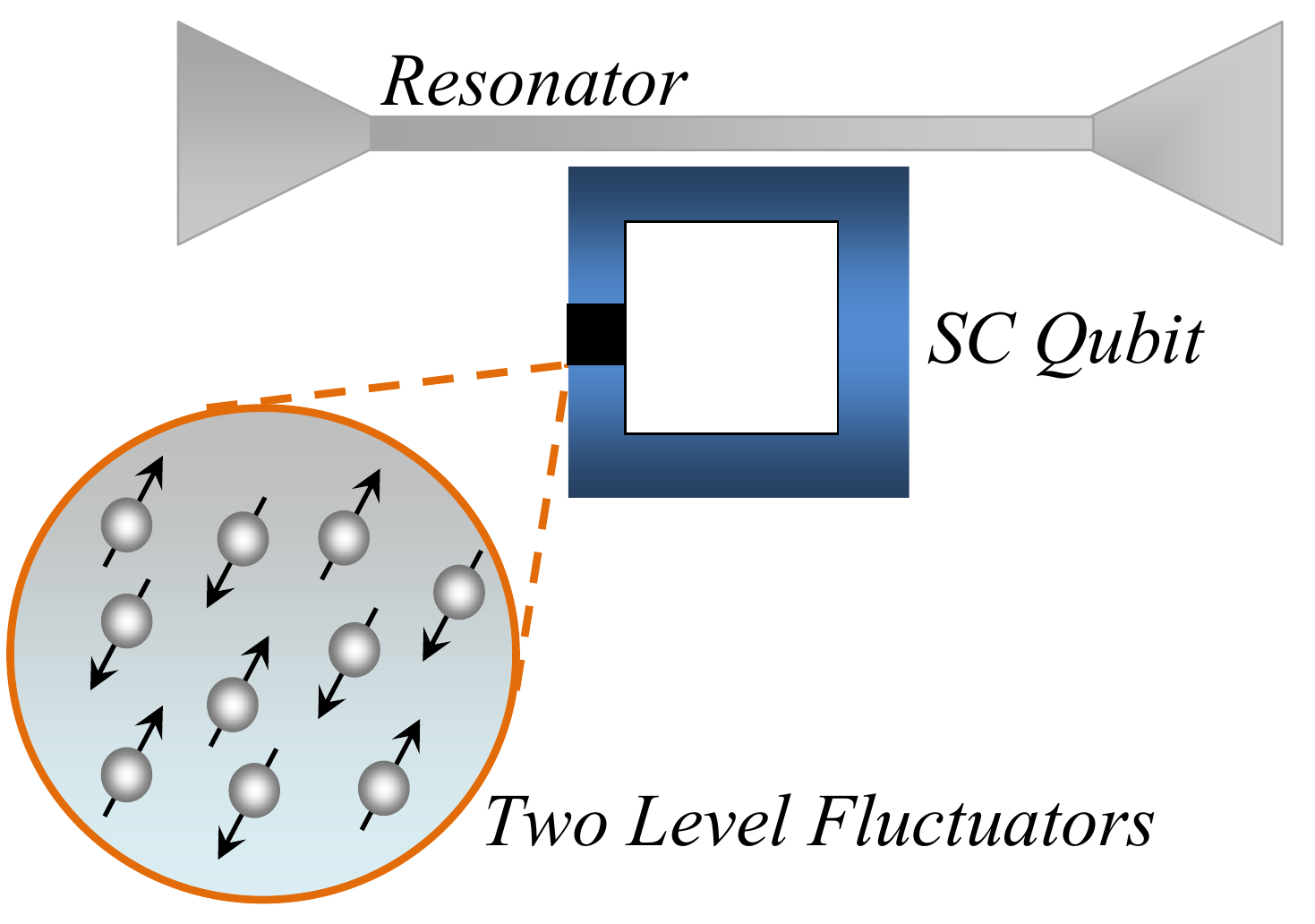}
\caption{ Schematic showing a superconducting(SC) qubit, coupled to a resonator, under the influence of several fluctuating two level systems present in the SQUID's metal-insulator interface.}
\label{fig:sch}
\end{figure}

Relatively recent experiments\cite{Sendelbach2008} suggest that flux noise arises from unpaired surface spins which reside at the superconductor-insulator interface in thin-film SQUIDs. Choi {\it et al}\cite{Choi2009} have explained this experimentally estimated areal spin density in terms of metal induced gap states that arise due to the potential disorder at the metal-insulator interface. These same set of experiments have observed strong correlation between the inductance noise and the flux noise, suggesting that the surface spins are interacting\cite{Sendelbach2008}. This appearance of a type of a long range magnetic order is somewhat consistant with a model suggested by Faoro and Ioffe \cite{Faoro2008} where the spins at the superconductor-insulator interface interact via the Ruderman–-Kittel–-Kasuya–-Yosida (RKKY) mechanism.

Regardless of the exact microscopic details of the noise mechanism, there appears to be a consensus that the $1/f$ type magnetization noise arises from the presence of TLSs. The TLSs generate random telegraph noise (RTN), which with a wide distribution of switching rates, gives $1/f$ noise\cite{Mottonen2006,Galperin2006} and behave very much as classical stochastic variables such as fluctuating Ising spins. A quasi-Hamiltonian formalism was introduced recently which was found to be extremely suitable and versatile for describing the non-unitary temporal evolution of a quantum system acted on by such a classical stochastic process \cite{Cheng2008,Joynt2009,Zhou2010}.

However, in a circuit QED type of a setup, the qubit is not only coupled to the TLSs, but also to a cavity mode. This further complicates the picture. In the Jaynes cummings model, the offdiagonal dipole coupling term between the superconducting qubit and cavity can be roughly expressed as $\lambda\propto\mu\cdot{B_{rms}}$\cite{Lindstrom2007}($\mu$ being the magnetic moment of the qubit and $B_{rms}$ is the rms value of photon's magnetic field). Now if a single TLS has a random time dependent fluctuating magnetic moment $m_j(t)$, then it can be argued that the dipole coupling term will now look like $\lambda\propto[\mu+m_j(t)]\cdot{B_{rms}}$. Thus fluctuations due to pure dephasing noise will induce fluctuations in the dipole coupling term which then introduces an energy relaxation mechanism.

The experimentally observed relevant energy scales are close enough for these effects to be noticeable. In case of a phase qubit, the vacuum Rabi splitting energy is about $\lambda=100$ MHz\cite{Poudel2012}(upto a factor of $\hbar$). While the estimated TLS's spliting energy is about $10$ MHz \cite{Yoshihara2006,Zhou2010} and has been reported to be even as high as $45$ MHz \cite{Yoni2010}.

In this paper, we extend a recently developed quasi Hamiltonian formalism\cite{Cheng2008,Joynt2009,Zhou2010} to treat a qubit-resonator system that is affected by the presence of multiple TLSs. This system schematically shown in Fig.\ref{fig:sch}. In the quasi-Hamiltonian method, the non-unitary temporal evolution of a quantum system acted on by a classical stochastic process in described in terms of the evolution of a non-Hermitian Hamiltonian. Decay processes in open quantum systems have been treated using using non-Hermitian Hamiltonians in the past \cite{Baker1984,Dattoli1990,deSouzaDutra2005,Fleischer2005,Rotter2009}.

Quite commonly, dissipative open quantum systems are treated using the Lindblad master equation\cite{Blum.book}. The Lindblad formalism can be extended to quantum optics type systems\cite{Carmichael.book,Kaer2010} and can be solved by the use of methods such the stochastic wavefunction method\cite{Dalibard1992,Imamoglu1994,Castin1995,Breuer1999} when many degrees of freedom are wanted. The stochastic wavefunction method has become a popular numerical technique since in comparison to the reduced density matrix method as it provides additional information about the state of the system\cite{Breuer1999}. However, this and other Monte-Carlo wave function methods \cite{Carmichael.book,Dalibard1992,Dum1992} are somewhat numerically intensive.

In our formalism, the dissipative dynamics of the entire qubit-resonator-fluctuator system is obtained by a {\it single} shot calculation. At the same time, we are able to individually set noise parameters for each fluctuator which enables us to introduce a broad distribution of relaxation rates (which lead to $1/f$ noise) and TLS splitting energies. Another advantage of our method is that temperature dependence and spin-spin interactions (such as RKKY for the fluctuators), can be easily introduced -- both of which are known to affect $1/f$ noise in SQUIDs \cite{Faoro2008,Bialczak2007}.


Or calculations shows that the system displays some unusual characteristics due to the stochastic fluctuations. For example, behavior that closely resembles collapse and revival of Rabi oscillations occur at times that depend on the noise coupling strength. A broad distribution of noise coupling strengths leads to a broad distribution of this type of behavior over time. It is also seen that, particularly for non-Markovian noise, the quantum photon bath can either speed up or slow down the relaxation and dephasing processes depending on the initial conditions.

This paper is organized as follows. In sec.\ref{sec.model} we discuss the model and our calculation method using the quasi Hamiltonian formalism within the Jaynes Cumming
s model. In sec.\ref{sec.Results} we discuss we present approximate analytic solutions for the case of a single TLS. The effect of mutiple TLS with fixed or random noise strength distribution and with a $1/f$ noise power spectrum is then discussed. This is followed by a particular example of a phase qubit in sec.\ref{subsec.pq}. Lastly we present our summary.

\section{Model and Methodology}\label{sec.model}

Consider the schematic of a superconducting(SC) qubit coupled to a resonator , as shown in fig.\ref{fig:sch}. At the metal insulator interface there exist a number of spins which randomly flip at different instances of time. The Hamiltonian describing the entire system is as follows

\begin{eqnarray}
H = H_s + H_{b} + H_{sb}
\label{H}
\end{eqnarray}

The system Hamiltonian, $H_s$ ,  for a single qubit (or a two level atom) in the presence of a single mode quantized electromagnetic field is given by the Rabi model. In the rotating wave approximation, or the Jaynes-Cummings model, only the energy conserving interaction term is retained
\begin{eqnarray}
H_s = \frac{\omega_o}{2}\sigma_z + \omega'{a^{\dagger}a} + \frac{\lambda}{2}(a\sigma_+ + a^\dagger\sigma_-)
\label{Hjc}
\end{eqnarray}
where, $\omega_o/2$ is the energy separation between the excited state $|1\rangle$, and the ground state, $|0\rangle$, of the qubit and is proportional to the applied magnetic field. Note that we have set $\hbar=1$. Here $\omega'$ is the photon frequency, $a^\dagger$ and $a$ are the photon energy creation and annihilation operators for a single mode, $\sigma_{\pm}=\sigma_x\pm i\sigma_y$. In its most general form, $\lambda\propto\langle{0}|\hat{d}|{1}\rangle$ is the dipole coupling term between the qubit states and $\hat d$ is the dipole operator. The exact form of $\lambda$ depends of the SC qubit type and resonator type.

The fluctuating two level systems at the metal insulator interface are modeled as flip-flopping Ising spins (fluctuators). If the fuctuators are statistically independent, then the noise Hamiltonian can be expressed in terms of the sum of contributions from individual fluctuations acting on the qubit
\begin{eqnarray}
H_b = \frac{1}{2}\displaystyle\sum_j{s_j(t){\bf g}^{(j)}\cdot\mathbf\sigma }
\label{RTN}
\end{eqnarray}
where, $s_j(t)$ is a two level sequence that switches between $\pm1$ at random intervals of time, ${\bf g}^{(j)}$ is the noise vector for the $j^{th}$ fluctuator and $\left\vert \mathbf{g}\right\vert =g$ is the noise strength. Their autocorrelation function is $\langle{s_i(t_1)s_j(t_2)}\rangle\propto\exp(-2\gamma_j|t_1-t_2|)\delta_{ij}$, $\gamma_j$ is the switching rate of the $j^{th}$ fluctuating Ising spin and each fluctuator has a Lorentzian power spectrum.

For our calculations we assume that the SC qubit's working point is such that, by it self, the SC qubit only has pure dephasing noise, {\it i.e.} ${\bf g_{\rm i}}=[0,0,g_{i}^{z}]$. The noise Hamiltonian then reduced to $H_b=\sum{s_i(t)g_i^z\sigma_z }$.

This noise Hamiltonian causes fluctuations in the energy levels of the qubit, which perturbs the wave function $|1\rangle\rightarrow|1\rangle'$. This in turn will cause fluctuations in the dipole coupling term $\lambda\propto\langle{0}|\hat{d}|{1}\rangle'$. This fluctuating dipole term will result in the presence of an off-diagonal energy relaxation term in the $2\times 2$ subspace of the Jaynes Cummings Hamiltonian.
\begin{eqnarray}
H_{sb} = \frac{1}{2}\displaystyle\sum_j g_x^{(j)} s_j(t) (a\sigma_+ + a^\dagger\sigma_-)
\label{Hjc}
\end{eqnarray}
where $g_x^{(j)} \propto \langle{0}|\hat{d}|{1}\rangle'-\lambda$ and has exactly the same time dependence as $g_z^{(j)}$. An alternate form of  $g_x^{(j)}$, more suitable for a phase/flux qubit, is discussed in sec. \ref{subsec.pq}.

More explicitly the full Hamiltonian  in the $\{|0;n+1\rangle,|1;n\rangle\}$ subspace is
\begin{equation}\small
H_n  = \displaystyle\sum_j\left[
\begin{array}{cc}
n\omega+ \frac{\omega_o}{2} + \frac{g_z^{(j)}}{2} s_j(t)   &  \left(\frac{\lambda + g_x^{(j)}}{2} s_i(t)\right)\sqrt{n+1} \\
\left(\frac{\lambda + g_x^{(j)}}{2} s_j(t)\right) \sqrt{n+1}  &  n\omega+ \frac{\omega_o}{2} + \Delta - \frac{g_z^{(j)}}{2}s_j(t)
\end{array}
\right]
\label{Hn}
\end{equation}
\noindent where $\Delta=\omega'-\omega_o$ is the field detuning.

$~~~$
\subsection{Transfer Matrix for the Qubit and the Photons}\label{sec.TM}

As we have mixed states, we seek solutions in terms of the density operator. We assume that at $t=0$, the qubit and the single cavity mode is not entangled and are pure states, {\it i.e.} the system's initial density matrix is $\rho(0)=\rho_Q(0)\otimes\rho_F(0)$. We can also further assume that the qubits are initially in the $|0\rangle$ state. Therefore, the initial density operators for the qubit and the photon field, respectively, are
\begin{eqnarray}
 \rho_Q(0)&=& |0\rangle \langle 0| \\
 \rho_F(0)&=& |\Psi_F(0)\rangle \langle\Psi_F(0)| = \displaystyle{\sum_n} |C_n|^2|n\rangle\langle n|\\\nonumber
\label{rho_qft}
\end{eqnarray}
The time dependent density matrix is
\begin{eqnarray}
\rho(dt)=U[\rho_Q(0)\otimes\rho_F(0)]U^{\dagger}
\label{rho_dt}
\end{eqnarray}
where $U=\exp(-iH dt)$. We next take a partial trace over the field to obtain the reduced density operator of the qubit
\begin{eqnarray}
\rho_Q(dt) &=& Tr_F[\rho(dt)] \\
           &=& \displaystyle\sum_n \langle n|U[\rho_Q(0)\otimes\rho_F(0)]U^{\dagger}|n\rangle.
\label{rho_dt_trF}
\end{eqnarray}

In general, the density operator for a single qubit can be written as
\begin{equation}
\rho (t)= \frac{1}{2}\left[I+\displaystyle\sum_{k=x,y,z} \eta_k(t)\sigma_k\right]
\label{rho-SU2}
\end{equation}
where, $\eta_k(t)$ are components of the Bloch vector, ${\mathbf\eta}(t)$, and $\left\vert{\mathbf\eta}\right\vert $ is a measure of purity.

The temporal dynamics of the quantum system can be reformulated as a transfer matrix equation by using the identity, given by Eq.\ref{rho-SU2}, on both sides of Eq. \ref{rho_dt_trF}
\begin{eqnarray}\small
I + \displaystyle\sum_k \eta_k(dt)\sigma_k
&=&   \sum_n\langle n| U (I\otimes\rho_F(0) U^{\dagger} |n\rangle \\\nonumber
&~& + \sum_{n,k} \langle n| U[ \eta_k(0)\sigma_k \otimes \rho_F(0)]U^{\dagger}|n\rangle.
\label{rho_dt_trF2}
\end{eqnarray}
Noting that $Tr[\rho_F(0)]=\sum|C_n|^2=1$, we obtain
\begin{eqnarray}
\displaystyle\sum_k \eta_k(dt)\sigma_k &=& \sum_{n,k} \langle n| U[ \eta_k(0)\sigma_k \otimes \rho_F(0)]U^{\dagger}|n\rangle~~~~~~~\\
                                        &=& \sum_{n,k} |C_n|^2 \eta_k(0) U_n \sigma_k U_n^{\dagger}
\label{rho_dt_trF4}
\end{eqnarray}
Also note that the unitary matrix, $U$, is block diagonal, where each of its $2\times2$ blocks are $U_n=e^{-iH_nt}$. Multiplying both sides of Eq.\ref{rho_dt_trF4} by $\sigma_j$ and using the identity $Tr[\sigma_k\sigma_j]=2\delta_{kj}$, the following transfer matrix equation can be obtained

\begin{eqnarray}
{\bf\eta}(dt)  = {\bf T}\cdot{\bf\eta}(0)
\label{TMeq}
\end{eqnarray}
where the elements of the $3\times 3$ transfer matrix $\bf{T}$ are
\begin{equation}
T_{kj}= \sum_n |C_n|^2 Tr\left[U_n\sigma_{k}U_n^{\dagger}\sigma_{j}\right]
\label{TM}
\end{equation}
As an example, in the presence of only a single TLS(with $s(t)=\pm1$), the two possible transfer matrices in the small time approximation are
\begin{eqnarray}
{\bf T}_\pm(dt)&=&\sum_n|C_n|^2 \{I + L_z(\Delta \mp g_z) dt ~~~~\\\nonumber
&~&~~~~~~~~~~~~+ L_x(\lambda \pm g_x)\sqrt{n+1} dt \}
\label{TM_expr}
\end{eqnarray}
Similarly, for $N$ TLS, there are $2^N$ possible transfer matrices ${\bf T}_{[\pm,\pm,...\pm]}$. 


\subsection{Quasi Hamiltonian in the Presence of Many TLS}\label{subsec.Gamma}


The overall temporal evolution of the quantum system (single cavity mode plus qubits) is governed by the time ordered product of the transfer matrices obtained in the previous section
\begin{equation}
\mathbf{\eta}(t)=\displaystyle\prod_{m=1}^{t/dt}\mathbf{T}_{m}\cdot \mathbf{\eta}(0).
\label{trans_t}
\end{equation}%

Whereas the classical stochastic noise process, for the two level fluctuators, is governed by the master equation, $\dot{\mathbf{W}}(t)=\mathbf{VW{\rm\it(t)}}$ \cite{VanKampen.book}, where $\mathbf{V}$ is a matrix of transition rates (such that the sum of each of its columns is zero) and $\mathbf{W}$ is the flipping probability matrix for the two level systems. For $N$ {\it uncorrelated} fluctuators, the flipping probability matrix is:
\begin{equation}
\mathbf{W}=\mathbf{W}_1\otimes\mathbf{W}_2\otimes...\mathbf{W}_N
\label{W}
\end{equation}
where the flipping probability matrix for a single fluctuator, assuming equal occupation probability of the two states, is
\begin{equation}
\mathbf{W_j}(t)=\frac{1}{2}\left[
\begin{array}{cc}
1+e^{-2\gamma_j t} & 1-e^{-2\gamma_j t} \\
1-e^{-2\gamma_j t} & 1+e^{-2\gamma_j t}%
\end{array}%
.\right]   \label{Wj}
\end{equation}
Here $\gamma_j$ is the rate at which the fluctuator $s_j(t)$ switches between $\pm1$.

In a manner similar to path integrals, one can therefore construct a combined transfer matrix that describes the small time($dt$) evolution of the quantum system coupled to the stochastic system (from the two types of transfer matrices discussed here) as follows
\begin{equation}
\mathbf{\Gamma }_{j}=\mathbf{W}\odot\mathbb{T}.
\label{Gamma}
\end{equation}
Where $\odot$ denotes a Hadamard product and $\mathbb{T}$ is a square matrix ($n$ being the number of qubits), each of whose columns consists of the lexicographically ordered transfer matrices $[{\bf T}_{[+,+,..+]}, {\bf T}_{[+,+,..-]} ... {\bf T}_{[-,-,..-]}]$ (the subscripts denote the TLS configuration).


The state of the quantum system along with $N$ classical TLSs ,at time $t$, is given by $\mathbf{\Gamma }^{m}=\mathbf{\Gamma }_{m}\mathbf{\Gamma }_{m-1}...\mathbf{\Gamma }_{1}$, where $m=t/dt$. By equating $\Gamma^m=\exp(-iH_qt)$, one can obtain $H_{q}=\displaystyle\lim_{dt\rightarrow 0}i(\mathbf{\Gamma -I})/dt$ in the small time limit. Where $H_{q}$ is a time-independent non-Hermitian quasi-Hamiltonian\cite{Joynt2009}. Its non Hermiticity implies that the time evolution operator $\exp(-iH_qt)$ need not be unitary, and hence can be used to treat dissipation in open quantum systems.

Finally we arrive at our main set of tools for calculating dissipation in this system. In the presence of $N$ fluctuating TLSs and a single cavity mode, the quasi Hamiltonian is
\begin{eqnarray}
\hat{H}^{(n)}_q &=& \hat{H}_{q\gamma} + \hat{H}_{qg} + \hat{H}_{q\omega}
\label{Hq_Nf}
\end{eqnarray}
\noindent where
{\begin{eqnarray}
\hat{H}_{q\gamma} &=& i\displaystyle\sum_{j=1}^N\gamma_j(\tau_x^{(j)}-I)\otimes{L_0} ~~~\\
\hat{H}_{qg}&=&i\displaystyle\sum_{j=1}^{N}  \tau_z^{(j)}\otimes (g_z^{(j)}L_z+g_x^{(j)}\sqrt{n+1}L_x)\\
\hat{H}_{q\omega} &=& iI\otimes(\Delta L_z -\lambda\sqrt{n+1}L_x).~~~~
\label{Hq_comp}
\end{eqnarray}}
\noindent Here, $L_{i=x,y,z}$ are the $SO(3)$-generators, $L_0$ is a $3\times3$ identity matrix, $I$ is a $2^N\times2^N$ identity matrix and it is implied that
\begin{eqnarray}
\tau_{x(z)}^{(j)} = \sigma_0^{(1)}\otimes...\sigma_0^{(j-1)}\otimes\sigma_{x(z)}^{(j)}\otimes\sigma_0^{(j+1)}\otimes...\sigma_0^{(n)}.~~~~~
\label{tau}
\end{eqnarray}

Finally, the time dependent Bloch vector for the qubit is obtained using the following projection
\begin{equation}\small
{\mathbf\eta}(t)=\langle f_N|...\langle f_2|\otimes\langle f_1|\sum_n|C_n|^2e^{-iH^{(n)}_{q}t}|i_1\rangle\otimes|i_2\rangle...|i_N\rangle{\mathbf\eta}(0).
\label{THqn}
\end{equation}
where $|i\rangle $ and $|f\rangle$ are the initial and final state vectors for a single TLS, that satisfy $\mathbf{W}|i(f)\rangle =|i(f)\rangle $ (these correspond to the zero eigenvalue solution of $\mathbf{V}$). For an unbiased TLS ({\it i.e.} with equal occupation probabilities), $|i\rangle =|f\rangle =[1,1]/\sqrt{2}$.


\section{Results and Discussion}\label{sec.Results}

        \subsection{Single Fluctuator}\label{subsec.sf}

We begin our discussion by considering a qubit coupled to a resonator under the influence of a single fluctuator. For a single cavity mode initially in a coherent state,
\begin{eqnarray}
|C_n|^2=\exp(-\langle n \rangle)\frac{{\langle n \rangle}^{n}}{n!}
\label{Cn_cohr}
\end{eqnarray}
where $\langle{n}\rangle$ is the average photon number and $|C_n|^2$ is the probability that there are $n$ photons present at $t=0$.

In order to obtain a clearer understanding of the Bloch vector's dissipative dynamics in such an environment, we solve Eq.\ref{THqn} numerically and obtain analytical expressions as well. The matrix exponential in the time evolution operator can be approximated using the Zassenhaus expansion \cite{Suzuki1977} as follows
\begin{eqnarray}
e^{-iH_qt}\approx e^{-iH_q't}e^{-iH_{q\omega}t} e^{[H_q',H_{q\omega}]t^2/2}
\label{eHq}
\end{eqnarray}
where $H_q'=H_{q\gamma}+H_{qg}$ and the commutator $ [H_q',H_{q\omega}] = -(g_z\lambda + g_x\Delta)\sigma_z\otimes{L_y}$. If we further assume that the resonator and the qubit are in resonance ($\Delta=0$), and noting that Eq.\ref{eHq} is only valid at short times, $t$,  or when $g_z\lambda$ is small.

We obtain approximate analytic expressions for the Bloch vector components for two different sets of initial conditions. If the Bloch vector initially pointing in the $x$-direction, ({\it i.e.} $\eta_o=[1,0,0]$), then the following components are obtained
\begin{eqnarray}
\eta_x &=& \displaystyle\sum_{n=0}^{\infty}|C_n|^2\Big[ \cos(\lambda_n g_z t^2) \frac{g_{xn}^2  + g_z^2\zeta(t)}{g_{xn}^2+g_z^2}\\\nonumber
&~&+g_z\sin(\lambda_n g_z t^2)\sin(2\lambda_n t)\frac{\sin(\Omega t)}{\Omega}e^{-\gamma t} \Big]
\label{nx1}
\end{eqnarray}
\begin{eqnarray}
\eta_y = \displaystyle\sum_{n=0}^{\infty}|C_n|^2g_{xn}\sin(\lambda_n g_z t^2)\cos(2\lambda_n t)\frac{\sin(\Omega t)}{\Omega}e^{-\gamma t}~~
\label{ny1}
\end{eqnarray}
\begin{eqnarray}
\eta_z &=& \displaystyle\sum_{n=0}^{\infty}|C_n|^2\Big[ g_{xn}g_z\cos(\lambda_n g_z t^2)\frac{1 -\zeta(t)}{g_{xn}^2 + g_z^2}\\\nonumber
&~&-g_{xn}\sin(\lambda_n g_z t^2)\sin(2\lambda_n t)\frac{\sin(\Omega t)}{\Omega}e^{-\gamma t} \Big]
\label{nz1}
\end{eqnarray}
where
\begin{eqnarray}
\zeta(t)= \left[\cos(\Omega t) + \frac{\gamma}{\Omega}\sin(\Omega t)\right]e^{-\gamma t}
\label{zeta}
\end{eqnarray}
and $\Omega= \sqrt{g_{xn}^2 + g_z^2 - \gamma^2}$, $g_{xn}=g_x\sqrt{1+n}$ and $\lambda_n=\frac{1}{2}\lambda\sqrt{1+n}$.

Whereas if we start with $\eta_o=[0,0,1]$, then the following components are obtained
\begin{eqnarray}
\eta_x' = \displaystyle\sum_{n=0}^{\infty}|C_n|^2 g_{xn}g_z\cos(2\lambda_n t)\cos(\lambda_n g_z t^2) \frac{1 - \zeta(t)}{g_{xn}^2+g_z^2}~~
\label{nx2}
\end{eqnarray}
\begin{eqnarray}
\eta_y'&=& \displaystyle\sum_{n=0}^{\infty}|C_n|^2\Big[ \sin(2\lambda_n t)\cos(\lambda_n g_z t^2)\zeta(t)\\\nonumber
&~&+g_{z}\sin(\lambda_n g_z t^2)\frac{\sin(\Omega t)}{\Omega}e^{-\gamma t} \Big]
\label{ny2}
\end{eqnarray}
\begin{eqnarray}
\eta_z' = \displaystyle\sum_{n=0}^{\infty}|C_n|^2 \cos(2\lambda_n t)\cos(\lambda_n g_z t^2) \frac{g_z^2 + g_{xn}^2\zeta(t)}{g_{xn}^2+g_z^2}~~
\label{nz2}
\end{eqnarray}



Next, consider the Bloch vector's dissipative dynamics for two extreme cases. {\it Note}, that the Bloch vector dynamics shown in all the figures are obtained by numerically solving Eq.\ref{THqn}.

In the first case let the qubit be subjected only to pure dephasing random telegraphic noise from a single fluctuator with negligible coupling to the resonator. Under these conditions, if the Bloch vector is initialized along the $z$-direction, then it will remain there indefinitely (as seen in Eqs.\ref{nx2}-.\ref{nz2}, where $\eta_z'(t)=1$ and $\eta_x'(t)=\eta_y'(t)=0$, if $g_{xn}=\lambda_n=0$).

\begin{figure}
\includegraphics[width=1\columnwidth]{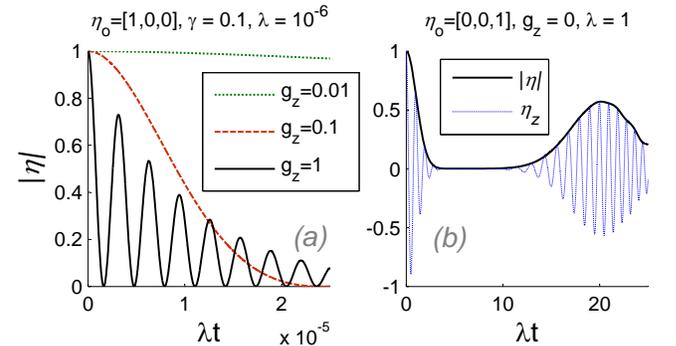}
\caption{Two extreme cases of the Bloch vector's dissipative dynamics. (a) TLS induced decoherence in the Markovian ($\gamma>g_z$), intermediate and non-Markovian noise regimes with negligible coupling to the resonator. (b) Collapse and revival type behavior of Rabi oscillations induced by $\langle{n}\rangle=10$ photons in the coherent state and in the absence of coupling to the TLS. Note the different initial conditions ($\eta_o$) in each case.}
\label{fig:nvst1}
\end{figure}

In contrast if the Bloch vector is initialized along the $x$-axis ($\eta_o=[1,0,0]$) then it will undergo either an oscillatory type decay or a monotonic decay to the center to the Bloch sphere as shown Fig.\ref{fig:nvst1}-a. From Eqs.\ref{nx1}-\ref{nz1}, it apparent that $\eta_x(t)$ is the only non-zero term if $g_{xn}=\lambda_n=0$.

In Fig.\ref{fig:nvst1}a, the decoherence of the Bloch vector is shown in three different noise coupling regimes -- in the strong coupling limit (non-Markovian noise, $g_z>\gamma $), in the intermediate regime ($g_z\approx \gamma )$ and in the weak coupling limit (Markovian noise $g_z<\gamma $). $\left\vert{\eta}\right\vert$ is a measure of the purity of the state and is zero only at the center of the Bloch sphere.

As one crosses over from the non-Markovian to the Markovian noise regime, the Bloch vector's decay goes from being oscillatory to be monotonic because the trigonometric functions in $\zeta(t)$ (see Eq.\ref{zeta}) become hyperbolic functions for Markovian noise ($\gamma >g_z+g_{xn}$). For non-Markovian noise, random Bloch vectors oscillations are caused by the $g_{z}s(t)\sigma _{z}$ term. For a given noise realization, the randomly switching $s(t)$ causes rotations about the $z$-axis, however, as ensemble averaging restores chiral symmetry, the averaged Bloch vector always travels in a straight line from pole to pole on the Bloch sphere\cite{De2011} (where the poles are $\vert+\rangle=(\vert{1}\rangle+\vert{0}\rangle)/\sqrt{2}$ and $\vert-\rangle=(\vert{1}\rangle-\vert{0}\rangle)/\sqrt{2}$) before diminishing to the center.

 Next, consider the case where the qubit is strongly coupled to the resonator and is completely decoupled from the fluctuator. The single cavity mode is initially prepared in a coherent state. In this case if $\eta_o=[1,0,0]$, then the Bloch vector will remain oriented along the x-direction indefinitely (which is again verifiable from Eqs.\ref{nx1}-\ref{nz1}). Whereas if the Bloch vector is initialized along the z-direction then it will undergo the well known phenomenon of collapse and revival(CR) of Rabi oscillations as shown in Fig.\ref{fig:nvst1}b. This CR process continues indefinitely with each revival being smaller in amplitude and less distinct from the preceding collapse. Increasing $\langle{n}\rangle$ results in more rapid collapses and more distinct revivals that are spaced further apart in time.

 These behavior can be explained as follows. In the limit of vanishing $g_z$ and $g_{xn}$, Eq.\ref{nz2} reduces to the well known form: $\eta_z'(t)=\sum|C_n|^2 \cos(2\lambda_n t)$. Each term in the summation (over $n$) represents Rabi oscillations (weighted by $|C_n|^2$) that are associated with a definite value of $n$. At $t=0$ these different Rabi oscillation terms are all correlated. As time increases the destructive interference between these weighted oscillatory terms leads to the collapse of Rabi oscillations. However, at longer times the $|C_n|^2\cos(2\lambda_nt)$ type terms will be in phase again and their constructive interference leads the the revival of Rabi oscillations. The revival phenomenon is of course is a purely quantum mechanical feature and is dependent on the photon distribution $C_n$. If one were to take a continuous distribution of photons, then no revival occurs after the collapse and the quantum dissipative process then becomes indistinguishable from that due to a classical stochastic process.

\begin{figure}
\includegraphics[width=0.95\columnwidth]{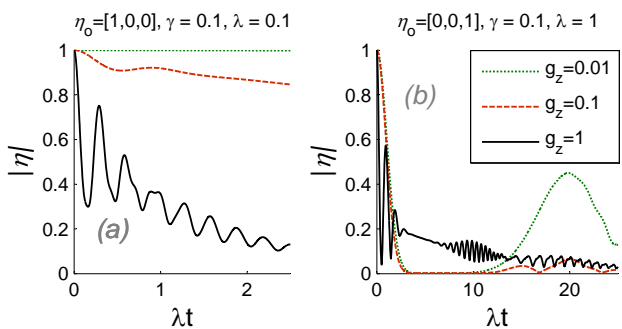}
\caption{The Bloch vector's decoherence in the Markovian ($\gamma>g_z$), intermediate and non-Markovian noise regimes with $\langle{n}\rangle=10$ photons in the coherent state for the Bloch vector initially in (a) $\eta_o=[1,0,0]$ with $\lambda=0.1$ and (b) $\eta_o=[1,0,0]$, $\lambda=1$. These calculations are carried out numerically.}
\label{fig:nvst2}
\end{figure}

Next, we consider the case where the qubit is subjected to a single fluctuator and is also coupled to the resonator with $\langle{n}\rangle=10$. As discussed in sec.\ref{sec.model}, in the presence of a pure dephasing noise, the fluctuating dipole coupling term($\lambda$) induces energy relaxation with noise strength $g_{xn}$. For our calculations we take $g_{xn}=g_z\lambda_n\sqrt(n+1)$.

The presence of $g_{xn}$ makes the picture more complicated as seen in Fig.\ref{fig:nvst2}-a, where the decoherence of the Bloch vector is shown in non-Markovian, intermediate and Markovian noise regimes for $\eta_o=[1,0,0]$. It is seen that the decoherence process is significantly slowed down for Markovian noise and noise in the intermediate coupling regime. Whereas for non-Markovian noise, this slowing down of the decoherence is less significant.
When the initial conditions are changed to $\eta_o=[1,0,0]$, then there is no slowing down of decoherence as seen in Fig.\ref{fig:nvst2}-b. On the contrary the coherence times terms deteriorate with increasing noise coupling strength, $g_z$. However, certain unusual collapse and revival type phenomenon occurs in the non-Markovian noise regime. While typically the occurrence of revivals (as a function of $\lambda{t}$) depends only on the photon distribution $|C_n|^2$, in the presence of a fluctuator, smaller revivals can occur at earlier times depending on $g_z$! As explained earlier, typically constructive interference between $\cos(2\lambda_nt)$ type terms (as permitted by $C_n$) leads to revivals for a pure resonator-qubit system. However the presence of the $\cos(g_z\lambda t^2)$ type terms in the expansion of the Bloch vector components (Eqs. \ref{nx2}-\ref{nz2}), will cause the constructive interference and hence the revival type phenomenon to occur earlier and with greater prominence, if $g_z$ is large (or for non-Markovian noise).


An experimental observation of this phenomenon would be a good indication that TLS are present.  If there are only a small number of TLS, then then a fit of the theory of this section to the results can reveal the coupling strengths and switching rates of the TLS.

\subsection{Multiple Fluctuators}\label{subsec.mf}

Typically, most experimental systems have many TLS with a broad distribution of switching rates.  In this section we extend our calculations to this case.

\begin{figure}
\includegraphics[width=1\columnwidth]{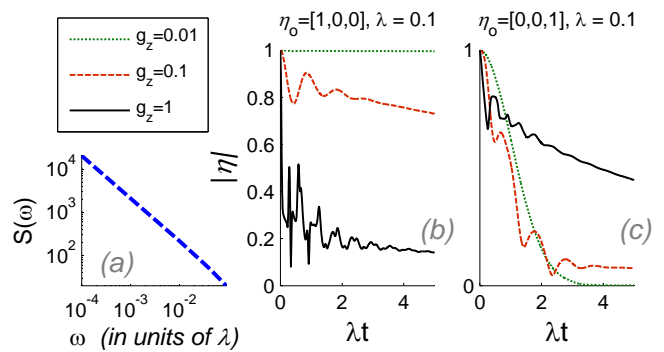}
\caption{ (a) Power spectrum showing 1/f noise for a set of eight TLSs. (b) Decoherence of the Bloch vector for different noise strengths with $\lambda=0.1$, $\langle{n}\rangle=10$ coherent state photons for an initial state $\eta_o=[1,0,0]$ and (c) $\eta_o=[0,0,1]$.}
\label{fig:BV_L01}
\end{figure}

\begin{figure}
\includegraphics[width=0.95\columnwidth]{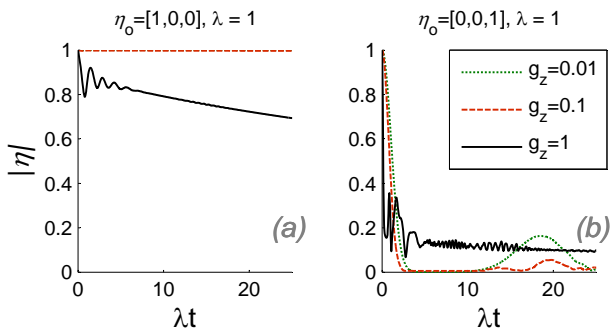}
\caption{(a) Decoherence of the Bloch vector for different noise strengths with $\lambda=1$, $\langle{n}\rangle=10$ coherent state photons for an initial state $\eta_o=[1,0,0]$ and (b) $\eta_o=[0,0,1]$. Note that the power spectrum for these fluctuators is the same as that shown in Fig.\ref{fig:BV_L01}-a.}
\label{fig:BV_L1}
\end{figure}

For the following set of calculations we use a set of eight fluctuators (as $\hat{H}_q^{(n)}$ scales as $2^n$, calculations with a large number of fluctuators quickly becomes prohibitive). Each individual fluctuator has a Lorentzian power spectrum and a broad distribution of their relaxation rates, $\gamma$ (ranging from $0.14$ to $2\times{10}^{-5}$), results in a $1/f$ noise power spectrum as shown in Fig.\ref{fig:BV_L01}-a. For simplicity and to limit the number of parameters used (in this section), we have held $g_j$s the same for all the fluctuators.

The decoherence of the Bloch vector for different noise strengths is shown in Fig. \ref{fig:BV_L01}-b and c for two different sets of initial conditions, $\eta_o$. As expected the overall rate of decoherence increases with increasing noise strength, $g_j$. The variations in the dissipation rate for different $g_j$ is most apparent if one starts with an initial state of $\eta_x$ as shown in Fig. \ref{fig:BV_L01}-b. For a full spectrum of fluctuators, if $g_j$ is smaller than the smallest $\gamma$ then the noise due to the TLSs falls purely in the Markovian noise regime, where the rate of dissipation will be the least. If $g_j$ falls somewhere in between the selected range of $\gamma$s, then one has a mixture of Markovian and non-Markovian noise sources, which is the case for $g_j=0.01$. However in this case, the Markovian noise sources tend to dominate and the oscillatory behavior (as apparent from Eq.\ref{zeta}), typically seen for non-Markovian noise, tends to get washed out, as shown for an initial state of $\eta_z$ (Fig. \ref{fig:BV_L01}-c). However now, collapse and revival of Rabi oscillations are seen at longer times due to the presence of a single photon mode, similar to that of Fig.\ref{fig:nvst2}-b (this is not shown in Fig. \ref{fig:BV_L01}, as we wish to focus on the initial collapse).

In the case of $g_j=0.1$ one has a mixture of Markovian and mostly intermediate noise sources. This leads to an oscillatory type of behavior where small oscillations are superposed on top of a smoothly decaying function. Now, if $g_j=1$, then one is entirely in the non-Markovian noise regime, and the decay of the Bloch vector is strongly oscillatory. However, now the decay of the Bloch vector is far more strongly affected by the resonator due to $\cos(g_z\lambda t^2)$ type terms in Eqs. \ref{nx1}-\ref{nz2}. The initial collapse of Rabi oscillations now mixes with TLSs induced oscillations with characteristic frequencies of $\Omega_j$. A broad distribution of $\Omega_j$ (due to $\gamma_j$) results in the complicated beating of the Bloch vector seen in Fig.\ref{fig:BV_L01} (for $g_j=1$). However now, the subsequent revival of Rabi oscillations (similar to that of Fig.\ref{fig:nvst2}-b for $g_j=1$) will not be seen at longer times as this behavior will be suppressed by $e^{-\sum\gamma_jt}$ due to the presence of multiple fluctuators. For this to be visible we have to increase the resonator coupling strength, which is what is done for the next set of calculations.

Fig.\ref{fig:BV_L1} shows the dissipation of the Bloch vector for $\lambda=1$ for two sets of initial conditions. It is apparent that on increasing $\lambda$, strong suppression of decoherence is seen in the Markovian and intermediate noise regimes for $\eta_o=\eta_x$ (see Fig.\ref{fig:BV_L1}-a). For non-Markovian noise the decoherence is more apparent, but still slower than the previous case of $\lambda=0.1$. For $\eta_o=\eta_z$, however, the dissipative behavior of the Bloch vector shows much richer behavior (see Fig.\ref{fig:BV_L1}-b). The unusual collapse and revival type phenomenon seen earlier for the single fluctuator case (see Fig.\ref{fig:nvst2}-b) occurs here as well for $g_j=1$. However for this to be more apparent in the case of multiple fluctuators, the noise must be strongly non-Markovian and the qubit should couple to the resonator strongly. Overall, $\lambda$ should be large enough to overcome the suppression of this collapse and revival type phenomenon by $e^{-\sum\gamma_jt}$.

\subsection{A Phase Qubit Example}\label{subsec.pq}


\begin{figure}
\includegraphics[width=1\columnwidth]{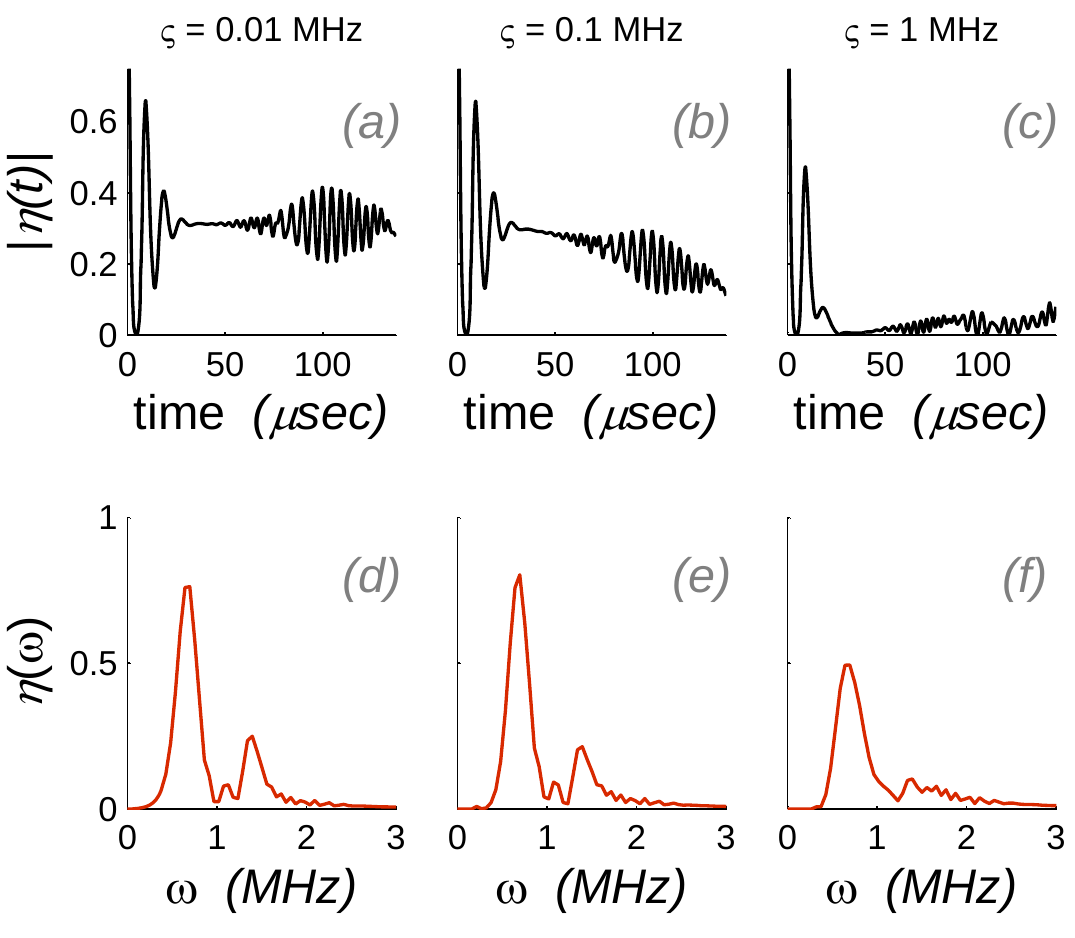}
\caption{Decoherence of the Bloch vector for $\lambda=100$ MHz,$g_{xo}=10^{-2}$, $\langle{n}\rangle=10$, $\eta_o=[0,0,1]$, and a normal distribution of $g_j$ with mean $\langle g\rangle=10$ MHz and standard deviation of (a) $\varsigma=0.001$ MHz, (b) $\varsigma=0.01$ MHz and (c) $\varsigma=0.1$ MHz.}
\label{fig:Rabi_FFT}
\end{figure}

\begin{figure}
\includegraphics[width=0.75\columnwidth]{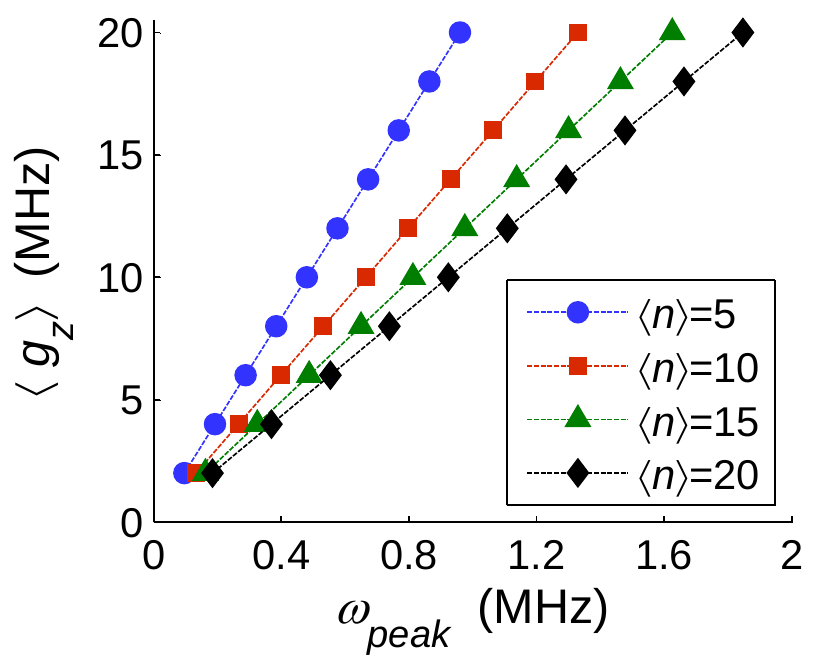}
\caption{Mean value, $\langle{g_z}\rangle$, as a function of the location of the peak values in the fourier spectrum (see Fig.\ref{fig:Rabi_FFT} d-f) frequency for various $\langle{n}\rangle$. These calculations were carried out for seven TLS with $\varsigma=0.1$ MHz and $\kappa=0.01$.}
\label{fig:g_vs_w}
\end{figure}

Our discussion of the TLS induced Bloch vector dynamics, in the previous section, was general and did not pertain to any particular type of qubit. In this section we consider the example of a phase qubit. In general, all superconducting qubits are characterized by the nonlinearity of its Josephson junctions, which can be thought of as a nonlinear inductor. When coupled to a capacitor (either external or arising from the junction), it forms a nonlinear LC oscillator. In the case of phase qubits, it so happens that the characteristic impedance of this nonlinear LC oscillator matches that of microwave transmission lines/resonators, which then allows for strong qubit-photon coupling.

The general Hamiltonian of a currant biased phase qubit is $H_{\varphi}=q^2/2C-I_o\Phi_o(\cos\varphi-I_b\varphi/I_o)/2\pi$. Where $q$ is the charge operator, $C$ is the capacitance, $\Phi_o$ is the magnetic flux quantum, $I_b$ is the bias currant of the Josephson junction(JJ), $I_o$ is its critical value and $\varphi$ is the phase across the JJ. By suitably biasing the JJ one obtains a washboard type potential energy landscape where the qubit states correspond to the two lowest quantized energy levels. This energy difference corresponds to $\omega_o$ in Eq.\ref{H}. In our calculations, however, this only appears in the form of the level detuning $\Delta$, when calculating the dissipative dynamics (see Eq.\ref{Hq_comp}). For all of our calculations we take $\Delta=0$. And for the vacuum Rabi splitting energy, we use the experimental value of $\lambda=100$ MHz\cite{Poudel2012}(note that $\hbar=1$).

Experimentally it is seen that the TLS have a gaussian distribution of splitting energies with a relatively small variance\cite{Yoni2010} for most fluctuators. Their estimated mean fluctuator splitting energies were about $10$ MHz, however a few of the $g_j$s were as high as about $40$ MHz. Other estimates of $g_j$ obtained by fitting to spin echo data vary between about $9.6$ MHz\cite{Yoshihara2006,Zhou2010} to values as high as $135$ MHz \cite{Zhou2010,Kakuyanagi2007}. Though the various experimental estimates vary, it is quite clear that the TLS splitting energy is much higher than a factor of $g_L\mu_B$ (where $g_L$ is the Land\'e $g$-factor and $\mu_B$ is the Bhor magneton) for an electron spin.

In the previous section $g_j$ was assumed to be the same for all the fluctuators. Here we examine how the collapse and revival type behavior would be affected if one were to consider a number of fluctuators with a random distribution of $g_j$. For the next set of calculations we consider a set of six Gaussian distributed random TLSs with a mean value of $\langle{g_z}\rangle=10$ MHz and a standard deviation of $\vartheta=0.01,0.1$ and $1$ MHz.

As mentioned earlier, the vacuum Rabi frequency $\lambda\propto\mu\cdot{B_{rms}}$\cite{Lindstrom2007}(where $\mu$ is the magnetic moment of the qubit and $B_{rms}$ is the root-mean-square value of photon's magnetic field). In the presence of multiple TLS with magnetic moment $m_j$, we argue that the dipole coupling term will look like $\lambda\propto[\mu+\sum_j{m_j}S(t)]\cdot{B_{rms}}$. From this, the fluctuating coupling term can be written as $g_{xn}^{(j)}=\kappa g_{z}^{(j)}\sqrt{1+n}$, where $\kappa$ is phenomenological scaling constant and we set $\kappa=0.01$. Now, note that all of our calculations, the collapse and revival type behavior is invariant with respect to the $\kappa{t}$ time scale. Hence the noise induced Bloch vector dynamics will be observable even for the smallest $\kappa$.

The resultant temporal dynamics and the respective Fourier transforms are shown in Fig.\ref{fig:Rabi_FFT}. The first peak in the Fourier spectrum corresponds to the initial set of oscillations while the second less prominent peak is due to the revived secondary set of oscillations. The width of these peaks corresponds to the various frequency components of these oscillations. As seen in the figure, the collapse and revival type phenomenon is more apparent if $\varsigma$ is small. For a larger $\varsigma$, the visibility of this effect diminishes due to the superposition of the collapse and revival type oscillations with more widely varying frequencies. This {\it washing out} effect is also apparent in the Fourier spectra, particularly in the significant broadening and lowering of the secondary peak. However, we also see that this {\it washing out} effect ,with increasing $\varsigma$, can be countered to some extent by increasing $\langle{n}\rangle$.

The dependency of the frequency of these CR oscillations on $\langle{g_z}\rangle$ and on $\langle{n}\rangle$ is examined more closely nest. In Fig.\ref{fig:g_vs_w}, the frequency at which the peak in the Fourier spectra occurs, $\omega_{peak}$ , (corresponding to the first set of oscillations) is shown as a function of $\langle{g_z}\rangle$ and $\langle{n}\rangle$. As expected $\omega_{peak}$ varies linearly with $\langle{g_j}\rangle$ and as $\langle{n}\rangle$ is increased $\omega_{peak}$'s dependence on $\langle{g_z}\rangle$ becomes more discernable.

In general the dependence of $\omega_{peak}$ on $\langle{g_z}\rangle$ can be somewhat complicated. However, based on our analytic solutions(for e.g. see Eq.\ref{nz2}), we found the following relation to be approximately true.
\begin{eqnarray}
\omega_{peak}\approx 2\kappa\langle{g_z}\rangle\sqrt{1+\langle{n}\rangle}
\label{wpeak}
\end{eqnarray}
Even more encouragingly, this relation is {\it independent} of the number of TLSs (which was explicitly verified for upto seven particles). Furthermore the accuracy of Eq.\ref{wpeak} improves with increasing average number of coherent photons ${\langle{n}\rangle}$.

This has important consequences from an experimental point of view. This implies that the mean value of the TLS splitting energy could be determined by loading different number of photons into the resonator and observing the frequency shift in $\omega_{peak}$. The only detrimental factor to this approach would be a vary large standard deviation, $\varsigma$, in the distribution of $g_j$s, which would make the Fourier spectrum less discernable. This however could still lead to further physical insight into the system.

\section{Summary}\label{subsec.Results}
In summary we have calculated the dissipative dynamics for a superconducting qubit coupled to a resonator, where the qubit is affected by a number of fluctuating two level systems with a $1/f$ noise power spectrum. The decoherence of the Bloch vector is calculated using a quasi Hamiltonian formalism whithin the Jaynes Cummings model.

We argue that fluctuations that cause pure dephasing noise will also induce an energy relaxation mechanism once the qubit is coupled to the resonator. The simultaneous coupling of a qubit to a coherent photon bath and classical TLS is shown to result unusual dissipative behavior. It is seen that collapse and revival of Rabi oscillations occur at times that depend on the noise coupling strength. This collapse and revival type behavior is particularly strong in the non-Markovian noise regime. Depending on the initial orientation of the Bloch vector, the coherent quantum photon bath either speeds up or slows down decoherence, which is also particularly strong for non-Markovian noise.

A broad distribution of noise coupling strengths leads to the broadening of the collapse and revival type behavior, which makes it less apparent. The frequency of these Rabi oscillations depends on the mean noise strength. We find an approximate analytic relation for this find it to be independent of the number of fluctuators. From an experimental point of view, it is perceivable that with controllable $\lambda$ and $\langle{n}\rangle$ one can estimate the average noise strength and various characteristics of its frequency spectrum.

The authors wish to thank Robert McDermott for a number of useful discussions. This work is supported by DARPA-QuEst Grant No. MSN118850.

\bibliographystyle{apsrev}

\end{document}